\documentclass{vgtc}
\ifpdf%                                % if we use pdflatex
  \pdfoutput=1\relax                   % create PDFs from pdfLaTeX
  \usepackage{graphicx}                % allow us to embed graphics files
  \DeclareGraphicsExtensions{.pdf,.png,.jpg,.jpeg} % for pdflatex we expect .pdf, .png, or .jpg files
\else%                                 % else we use pure latex
  \usepackage{graphicx}                % allow us to embed graphics files
  \DeclareGraphicsExtensions{.eps}     % for pure latex we expect eps files
\fi%

\graphicspath{{figures/}{pictures/}{images/}{./}} % where to search for the images

\usepackage{microtype}                 % use micro-typography (slightly more compact, better to read)
\PassOptionsToPackage{warn}{textcomp}  % to address font issues with \textrightarrow
\usepackage{textcomp}                  % use better special symbols
\usepackage{mathptmx}                  % use matching math font
\usepackage{times}                     % we use Times as the main font
\usepackage{cite}                      % needed to automatically sort the references
\usepackage{tabu}                      % only used for the table example
\usepackage{booktabs}                  % only used for the table example
\usepackage{amsfonts,amsmath,amsthm}
\usepackage{array}
\usepackage{booktabs}
\usepackage{color}
\usepackage{enumitem}
\usepackage{float}
\usepackage{mathtools}
\usepackage{multirow}
\usepackage{setspace}

\theoremstyle{definition}

\renewcommand\footnotemark{}

\newcolumntype{L}[1]{>{\raggedright\let\newline\\\arraybackslash\hspace{0pt}}m{#1}}
\newcolumntype{C}[1]{>{\centering\let\newline\\\arraybackslash\hspace{0pt}}m{#1}}
\newcolumntype{R}[1]{>{\raggedleft\let\newline\\\arraybackslash\hspace{0pt}}m{#1}}
\setlist[itemize]{noitemsep, topsep=0pt}
\setlist[enumerate]{noitemsep, topsep=0pt}

\newcommand{\sota}{state-of-the-art}

\newcommand{\parens}[1]{\left(#1\right)}
\newcommand{\braces}[1]{\left\{#1\right\}}
\newcommand{\bracks}[1]{\left[#1\right]}

\onlineid{0}

\vgtccategory{Research}

\vgtcinsertpkg

\title{Text2Gestures: A Transformer-Based Network for Generating Emotive Body Gestures for Virtual Agents\thanks{This work has been supported in part by ARO Grants W911NF1910069 and W911NF1910315, and Intel. Code and additional materials available at: \url{https://gamma.umd.edu/t2g}.}
}

\author{Uttaran Bhattacharya$^1$ %
\and Nicholas Rewkowski$^2$ %
\and Abhishek Banerjee$^3$ %
\and Pooja Guhan$^4$ %
\and Aniket Bera$^5$ %
\and Dinesh Manocha$^6$ %
}
\affiliation{\scriptsize University of Maryland, College Park, MD 20742, USA \\ $^1$\url{uttaranb@umd.edu}\hfill
$^2$\url{nick1@umd.edu}\hfill
$^3$\url{abanerj8@terpmail.umd.edu}\hfill
$^4$\url{pguhan@umd.edu}\hfill
$^5$\url{bera@umd.edu}\hfill
$^6$\url{dmanocha@umd.edu}
}

\abstract{
We present Text2Gestures, a transformer-based learning method to interactively generate emotive full-body gestures for virtual agents aligned with natural language text inputs. Our method generates emotionally expressive gestures by utilizing the relevant biomechanical features for body expressions, also known as affective features. We also consider the intended task corresponding to the text and the target virtual agents' intended gender and handedness in our generation pipeline. We train and evaluate our network on the MPI Emotional Body Expressions Database and observe that our network produces state-of-the-art performance in generating gestures for virtual agents aligned with the text for narration or conversation. Our network can generate these gestures at interactive rates on a commodity GPU. We conduct a web-based user study and observe that around 91\% of participants indicated our generated gestures to be at least plausible on a five-point Likert Scale. The emotions perceived by the participants from the gestures are also strongly positively correlated with the corresponding intended emotions, with a minimum Pearson coefficient of 0.77 in the valence dimension.%
} % end of abstract

\CCScatlist{
  \CCScatTwelve{Computing methodologies}{Virtual reality}{}{};
  \CCScatTwelve{Computing methodologies}{Intelligent agents}{}{};
  \CCScatTwelve{Computer systems organization}{Neural networks}{}{};
}

\begin{document}


%% \firstsection{Introduction}

\maketitle

\section{Introduction}\label{sec:intro}
%for journal use above \firstsection{..} instead
As the world increasingly uses digital and virtual platforms for everyday communication and interactions, there is a heightened need to create highly realistic virtual agents endowed with social and emotional intelligence. Interactions between humans and virtual agents are being used to augment traditional human-human interactions in different applications, including online learning~\cite{online_learning1,online_learning2,online_learning3}, virtual interviewing and counseling~\cite{interviewing,simsensei}, virtual social interactions~\cite{virtual_social_interactions1,virtual_social_interactions2,virtual_social_interactions3,generalized_combinations}, and large-scale virtual worlds~\cite{fb_horizons}. Human-human interactions rely heavily on a combination of verbal communications (the text), inter-personal relationships between the people involved (the context), and more subtle non-verbal face and body expressions during communication (the subtext)~\cite{human_human1,human_human2}. While context is often established at the beginning of interactions, virtual agents in social VR applications need to align their text with their subtext throughout the interaction, thereby improving the human users' sense of presence in the virtual environment. Gesticulation is an integral component in subtext, where humans use patterns of movement for hands, arms, heads, and torsos to convey a wide range of intent, behaviors, and emotions~\cite{gestures_in_comm}. In this work, we investigate the problem of aligning emotionally expressive gestures with the text to generate virtual agents' actions that result in natural interactions with human users.

Current game engines and animation engines can generate human-like movements for virtual agents, including head poses, hand gestures, and torso movements~\cite{nsm,speech_flow_gestures}. However, aligning these movements with a virtual agent's associated speech or text transcript is more challenging. Traditional approaches such as hand-crafting animations or collecting and transferring context-specific gestures through rotoscoping or motion capture look natural~\cite{prob_speaker_style,rule_based_gestures}, but need to be manually designed for every new gesture. However, virtual agents performing live social interactions with humans in VR need to adapt their gestures to their words and current social context in real-time. As a result, prior approaches based on pre-generated animations or motion specifications are limited, and we need interactive methods to generate plausible gestures.

Existing approaches for interactive speech-aligned gesture generation learn mappings between speech signals and the generated gesture sequences~\cite{speech_RepL_gestures,speech_flow_gestures}. In contrast to these speech-based methods, our goal is to align the gestures directly with the natural language text transcripts. This eliminates the need to have speeches pre-recorded by humans or machines, which have a higher production cost. Prior works on generating gestures aligned with text~\cite{cospeech_gestures} have leveraged the well-known sequence-to-sequence modeling network, which is efficient at performing a variety of sequence-to-sequence prediction tasks. These methods have only considered arms and head motions and are limited to generating gestures with small variations in categorical emotions such as happy, angry, and sad.

However, as evidenced by works on adding emotions through facial and vocal expressions~\cite{expressive_face2,verbal_comm2,panel_bodily_expressed_emotions}, emotional expressiveness adds to the realism of virtual agents. Studies in psychology and affective computing show that body expressions also contain useful cues for perceived emotions~\cite{body_er,step,taew}, and often help disambiguate the emotions perceived from facial and vocal cues~\cite{misleading_cues,m3er,emoticon}. These body expressions are composed of biomechanical features known as \textit{affective features}. Common affective features include, among others, the rate of arm swings, stride lengths, shoulder and spine postures, and head jerks~\cite{affective_body_survey2}. More recent approaches for generating virtual agents with gait-based body expressions have leveraged the relevant gait-based affective features to improve the perceived naturalness of the animations~\cite{eva,fva,step}.

Following these works, we aim to generate body gestures for virtual agents in social VR settings to either narrate text-based content to human participants or continue a text-based conversation with human participants. We use affective features to make the gestures emotionally expressive, such that the human participants can perceive appropriate emotions from the virtual agents based on the natural language text.

\noindent\textbf{Main Results:} We present an end-to-end trainable generative network that produces emotive body gestures aligned with natural language text. We design our method for interactive applications, where a virtual agent narrates lines or takes part in a conversation. To this end, we make use of the transformer network~\cite{transformer}, and extend current approaches to work with gestures for virtual agents in 3D. We also adapt the gestures based on narration or conversation and the intended gender and handedness (dominance of left-hand or right-hand in gesticulation) of the virtual agents. We also make the gestures emotionally expressive by utilizing the relevant gesture-based affective features of the virtual agents.

To summarize, our contributions are four-fold:

\begin{itemize}
    \item A transformer-based network that interactively takes in text one sentence at a time and generates 3D pose sequences for virtual agents corresponding to gestures aligned with that text.
    \item Conditioning the generation process to follow the intended acting task of narration or conversation and the virtual agents' intended gender and handedness.
    \item Considering the intended emotion in the text to generate emotionally expressive gestures.
    \item A web study with 600 total responses to evaluate the quality of our generated gestures compared to motion-captured sequences and the emotional expressiveness of our generated gestures.
\end{itemize}

Based on our experiments, we find that our network has state-of-the-art performance for generating gestures aligned with text compared to ground-truth sequences in a large-scale motion capture database. We can generate these gestures at an interactive rate of 312.5 fps using an Nvidia GeForce GTX 1080Ti GPU. Based on our user study, we also find that the emotions perceived by the participants from the gestures are strongly positively correlated with the corresponding intended emotions of the gestures, with a minimum Pearson coefficient of 0.77 in the valence dimension. Moreover, around 91\% of participants found our generated gestures are plausible on a five-point Likert Scale.

\section{Related Work}\label{sec:rw}
This section summarizes studies exploring how different emotions are perceived from body gestures and how they have been utilized to generate emotive virtual agents.

We also review prior work on generating human body gestures in graphics and VR, particularly those that align the gestures with speech and text content. We focus mostly on data-driven approaches here because we base our work on a similar foundation, and refer the interested reader to Wagner et al.'s extensive survey~\cite{rule_based_gestures} for the more classical rule-based approaches. The main limitation of such rule-based approaches is that their range of gestures is confined to the designed set of gestures. Hence, they require that gestures for every novel speech and text inputs are manually designed.

\subsection{Perceiving Emotions from Body Expressions}
Studies in psychology show that body expressions, including gestures, are better suited than facial and vocal cues to express and perceive emotions varying in arousal and dominance, such as anger, relief, fear, and pride~\cite{neurobiology_of_emotional_body_expressions,effort_shape}. Body expressions are also useful for disambiguating between pairs of emotions such as fear or anger~\cite{rapid_emotional_body_language}, and fear or happiness~\cite{body_expressions_influence}. Follow-up studies in affective computing~\cite{affective_body_survey1,affective_body_survey2,emotion_in_gesture,learning_unseen_emotions} have identified sets of biomechanical features from body expressions, known as affective features, on which human observers focus when perceiving these different emotions from gestures. For example, rapid arm swings can indicate anger, an expanded upper body can indicate pride, and slouching shoulders can indicate fear or sadness.

In our work, we use such affective features observable from gestures to emote our generated virtual agents.

\subsection{Generating Emotive Virtual Agents}
Current approaches to endow virtual agents with emotional expressiveness make use of a number of modalities, including verbal communication~\cite{verbal_comm1,verbal_comm2}, face movements~\cite{expressive_face1,expressive_face2}, body gestures~\cite{bonding_in_conversations}, and gaits~\cite{eva}. In the context of generating emotional expressions aligned with speech, Chuah et al.~\cite{automated_emotive_agents} leveraged a dataset of words mapped to emotive facial expressions to generate virtual agents with basic emotions automatically. DeVault et al.~\cite{simsensei} developed a full-fledged virtual human counselor, using a pre-built corpus of mappings between mental states and body expressions to make their virtual agent appropriately expressive. In contrast to these approaches, we build a generalizable data-driven mapping to body gestures from a more diverse range of intended emotions associated with text transcripts, such that we can generate appropriately expressive gestures for out-of-dataset text sentences.

\subsection{Generating Gestures Aligned with Speech and Text}
There has been extensive deep-learning-based work on generating human body gestures that align with speech content in the recent past~\cite{dcnf}. Levine et al.~\cite{speech_hmm_gestures} used a hidden Markov model to learn latent mappings between speech and gestures. Hasegawa et al.~\cite{speech_LSTM_gestures} used recurrent neural networks to predict 3D pose sequences for gestures from input speech. More recently, Kucherenko et al.~\cite{speech_RepL_gestures} trained autoencoders to learn latent representations for the speech and the gesture data and then learned mappings between the two to generate gestures that are less sensitive to noise in the training data. By contrast, Alexanderson et al.~\cite{speech_flow_gestures} learned invertible sub transformations between speech and gesture spaces to stochastically generate a set of best-fitting gestures corresponding to the speech. Other approaches have also incorporated individual styles into gestures~\cite{individual_conversational_gestures}, added multiple adversarial losses to make the generated gestures look more realistic~\cite{multi_adversarial_gestures}, and even added prototypical rule-based behaviors such as head nods and hand waves based on the discourse~\cite{prototypical_behaviors}. These have culminated into works such as generating gestures for multiple speakers through style-transfer~\cite{style_transfer_multi_speaker}, and semantic-aware gesture generation from speech~\cite{gesticulator}.

Our approach is complementary to these approaches in that we learn mappings from the text transcripts of speech to gestures. This eliminates the noise in speech signals and helps us focus only on the relevant content and context. Learning from the text also enables us to focus on a broader range of gestures, including iconic, deictic, and metaphoric gestures~\cite{gestures_in_comm}. Our work is most closely related to that of Yoon et al.~\cite{cospeech_gestures}. They learn upper body gestures as PCA-based, low-dimensional pose features, corresponding to text transcripts from a dataset of TED-talk videos, then heuristically map these 3D gestures to an NAO robot. They have also followed up this work by generating upper-body gestures aligned with the three modalities of speech, text transcripts, and person identity~\cite{trimodal_gesture_generation}. On the other hand, we learn to map text transcripts to 3D pose sequences corresponding to semantic-aware, full-body gestures of more human-like virtual agents using an end-to-end trainable transformer network and blend in emotional expressiveness.

\subsection{Generating Stylistic Human Body Motions}
Generating speech- or text-aligned gestures with emotional expressiveness can be considered a sub-problem in generating stylistic human body motions, including facial motions, head motions, and locomotion. Existing approaches on face motions include generating lip movements and other face-muscle motions aligned with speech, using either recurrent neural networks~\cite{obamanet} or convolutional networks~\cite{expressive_face2}. Methods for generating head motions that convey the pace and intensity of speech have explored neural network architectures based on autoencoders~\cite{head_motion_ae} and generative adversarial networks~\cite{head_gan}. Methods to generate stylistic locomotion are based on convolutional networks~\cite{motion_editing}, parametric phase functions~\cite{pfnn}, and deeply learned phase functions~\cite{nsm} for different styles of walking. Recent approaches have also incorporated gait-based affective features to generate emotionally expressive walking~\cite{tanmay_emotions,step,gen_emo_gaits}. Moreover, there has been considerable progress in generating images and videos of body motions based on textual descriptions of moments and actions~\cite{video_from_text,text_guided_pose}.

In contrast, we aim to generate emotionally expressive gestures at interactive rates that correspond to text sentences. The space of gesture motions we explore is also different from the space of motions corresponding to locomotion, head motions, or facial muscle motions. Although there is some overlap with the space of head motions~\cite{head_motion_ae,head_gan}, the corresponding methods have not been extended to deal with full-body motions.

\section{Transforming Text to Gestures}\label{sec:method}
Given a natural language text sentence associated with an acting task of narration or conversation, an intended emotion, and attributes of the virtual agent, including gender and handedness, our goal is to generate the virtual agent's corresponding body gestures. In other words, we aim to generate a sequence of relative 3D joint rotations $\mathcal{Q}^*$ underlying the poses of a virtual agent, corresponding to a sequence of input words $\mathcal{W}$, and subject to the acting task $A$ and the intended emotion $E$ based on the text, and the gender $G$ and the handedness $H$ of the virtual agent. We therefore have
\begin{equation}
    \mathcal{Q}^* = \arg\max_\mathcal{Q} \textrm{Prob}\bracks{\mathcal{Q} | \mathcal{W}; A, E, G, H}.
    \label{eq:transduction}
\end{equation}

\subsection{Representing Text}\label{subsec:text_rep}
Following standard practices in NLP tasks, we represent the word at each position $s$ in the input sentence $\mathcal{W} = \begin{bmatrix} w_1 & \dots & w_s & \dots  & w_{T_\textrm{sen}}\end{bmatrix}$, with $T_\textrm{sen}$ being the maximum sentence length, using word embeddings $w_s \in \mathbb{R}^{300}$. We obtain the word embeddings using the GloVe model pre-trained on the Common Crawl corpus~\cite{glove}. We opt for GloVe based on our preliminary experiments, where it marginally outperformed other similar-dimensional embedding models such as Word2Vec~\cite{word2vec} and FastText~\cite{fasttext}, and had similar performance as much higher dimensional embedding models, \textit{e.g.}, BERT~\cite{bert}. We demarcate the start and the end of sentences using special start of sequence (SoS) and end of sequence (EoS) vectors that are pre-defined by GloVe.

\begin{figure}[t]
    \centering
    \includegraphics[width=0.7\columnwidth]{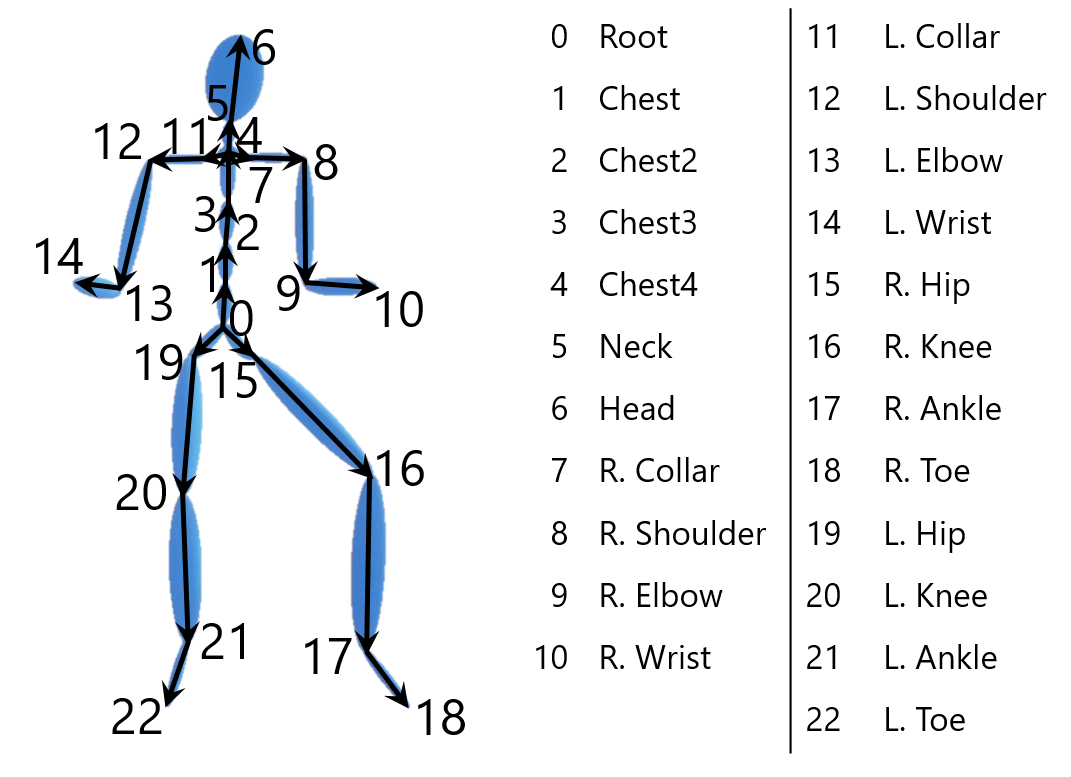}
    \caption{\textbf{Directed pose graph.} Our pose graph is a directed tree consisting of 23 joints, with the root joint as the root node of the tree, and the end-effector joints (head, wrists, toes) as the leaf nodes of the tree. We manipulate the appropriate joints to generate emotive gestures.}
    \label{fig:pose_graph}
    \vspace{-15pt}
\end{figure}

\subsection{Representing Gestures}\label{subsec:ges_rep}
Following prior works on human motion generation~\cite{quaternet}, we represent a gesture as a sequence of poses or configurations of the 3D body joints. These include body expressions as well as postures. We represent each pose with quaternions denoting 3D rotations of each joint relative to its parent in the directed pose graph (Fig.~\ref{fig:pose_graph}). Specifically, at each time step $t$ in the sequence $\mathcal{Q} = \begin{bmatrix} q_1 & \dots & q_t & \dots  & q_{T_\textrm{ges}}\end{bmatrix}$, with $T_\textrm{ges}$ being the maximum gesture length, we represent the pose using flattened vectors of unit quaternions $q_t = \begin{bmatrix} \dots & q_{j, t}^\top & \dots \end{bmatrix}^\top \in \mathbb{H}^J$. Each set of $4$ entries in the flattened vector $q_t$, represented as $q_{j, t}$, is the rotation on joint $j$ relative to its parent in the directed pose graph, and $J$ is the total number of joints. We choose quaternions over other representations to represent rotations as quaternions are free of the gimbal lock problem~\cite{quaternet}. To demarcate the start and the end of each gesture sequence, we define our start-of-sequence (SoS) and end-of-sequence (EoS) poses. Both of these are idle sitting poses with decorative changes in the positions of the end-effector joints, the root, wrists and the toes.

\subsection{Representing the Agent Attributes}\label{attr_rep}
We categorize the agent attributes into two types: attributes depending on the input text and attributes depending on the virtual agent.
% Emotive Gestures subsection

\subsubsection{Attributes Depending on Text}
In this work, we consider two attributes that depend on text, the acting task, and the intended emotion.

\paragraph{Acting Task}
We consider two acting tasks, narration and conversation. In narration, the agent narrates lines from a story to a listener. The gestures, in this case, are generally more exaggerated and theatrical. In conversation, the agent uses body gestures to supplement the words spoken in conversation with another agent or human. The gestures are subtler and more reserved. In our formulation, we represent the acting task as a two-dimensional one-hot vector $A \in \braces{0, 1}^2$, to denote either narration or conversation.

\paragraph{Intended Emotion}
We consider each text sentence to be associated with an intended emotion, given as a categorical emotion term such as joy, anger, sadness, pride, etc. While the same text sentence can be associated with multiple emotions in practice, in this work, we limit ourselves to sentences associated with only one emotion, owing primarily to the limitations in the dataset available for training. We use the NRC-VAD lexicon~\cite{nrc_vad} to transform these categorical emotions associated with the text to the VAD space. The VAD space~\cite{vad} is a well-known representation in affective computing to model emotions. It maps an emotion as a point in a three-dimensional space spanned by valence (V), arousal (A), and dominance (D). Valence is a measure of the pleasantness in the emotion (\textit{e.g.}, happy vs. sad), arousal is a measure of how active or excited the subject expressing the emotion is (\textit{e.g.}, angry vs. calm), and dominance is a measure of how much the subject expressing the emotion feels ``in control'' of their actions (\textit{e.g.}, proud vs. remorseful). Thus, in our formulation, the intended emotion $E \in \bracks{0, 1}^3$, where the values are coordinates in the normalized VAD space.

\subsubsection{Attributes Depending on the Agent}
We consider two attributes that depend on the agent to be animated, its gender $G$, and handedness $H$. In our work, gender $G \in \braces{0, 1}^2$ is limited to a one-hot representation denoting either female or male, and handedness $H \in \braces{0, 1}^2$ is a one-hot representation indicating whether the agent is left-hand dominant or right-hand dominant. Male and female agents typically have differences in body structures (\textit{e.g.}, shoulder-to-waist ratio, waist-to-hip ratio). Handedness determines which hand dominates, especially when gesticulating with one hand (\textit{e.g.}, beat gestures, deictic gestures). Each agent has exactly one assigned gender and one assigned handedness.

\subsection{Using the Transformer Network}\label{subsec:transformer}
Modeling the input text and output gestures as sequences shown in Secs.~\ref{subsec:text_rep} and~\ref{subsec:ges_rep}, the optimization in Eq.~\ref{eq:transduction} becomes a sequence transduction problem. We, therefore, approach this problem using a transformer-based network. We briefly revisit the transformer as originally introduced by Vaswani et al.~\cite{transformer}, and describe how we modify it for our transduction problem.

The transformer network follows the traditional encoder-decoder architecture for sequence-to-sequence modeling. However, instead of using sequential chains of recurrent memory networks, or the computationally expensive convolutional networks, the transformer uses a multi-head self-attention mechanism to model the dependencies between the elements at different temporal positions in the input and target sequences.

\begin{figure}[t]
    \centering
    \includegraphics[width=\columnwidth]{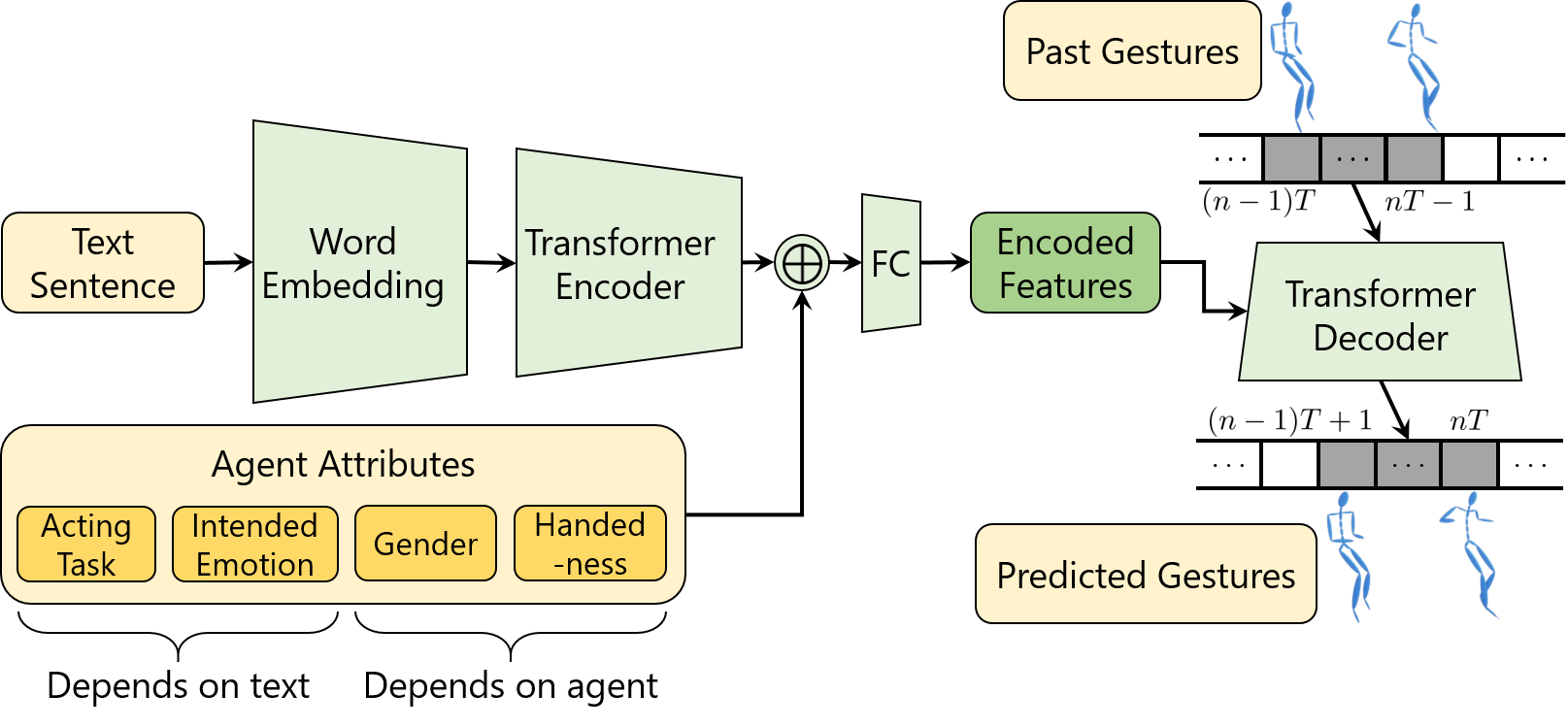}
    \caption{\textbf{Text2Gestures Network.} Our network takes in sentences of natural language text and transforms them to word embeddings using the pre-trained GloVe model~\cite{glove}. It then uses a transformer encoder to transform the word embeddings to latent representations, appends the agent attributes to these latent representations, and transforms the combined representations into encoded features. The transformer decoder takes in these encoded features and the past gesture history to predict gestures for the subsequent time steps. At each time step, we represent the gesture by the set of rotations on all the body joints relative to their respective parents in the pose graph at that time step.}
    \label{fig:network}
    \vspace{-15pt}
\end{figure}

The attention mechanism is represented as a sum of values from a dictionary of key-value pairs, where the weight or attention on each value is determined by the relevance of the corresponding key to a given query. Thus, given a set of $m$ queries $Q \in \mathbb{R}^{m \times k}$, a set of $n$ keys $K \in \mathbb{R}^{n \times k}$, and the corresponding set of $n$ values $V \in \mathbb{R}^{n \times v}$ (for some dimensions $k$ and $v$), and using the scaled dot-product as a measure of relevance, we can write,
\begin{equation}
    \textrm{Att}\parens{Q, K, V} = \textrm{softmax}\parens{\frac{QK^\top}{k}}V,
\end{equation}
where the softmax is used to normalize the weights. In the case of self-attention (SA) in the transformer, $Q$, $K$, and $V$ all come from the same sequence. In the transformer encoder, the self-attention operates on the input sequence $\mathcal{W}$. Since the attention mechanism does not respect the relative positions of the elements in the sequence, the transformer network uses a positional encoding scheme to signify the position of each element in the sequence, prior to using the attention. Also, in order to differentiate between the queries, keys, and values, it projects $\mathcal{W}$ into a common space using three independent fully-connected layers consisting of trainable parameters $W_{Q, enc}$, $W_{K, enc}$, and $W_{V, enc}$. Thus, we can write the self-attention in the encoder, $\textrm{SA}_{enc}$, as
\begin{equation}
    \textrm{SA}_{enc}\parens{\mathcal{W}} = \textrm{softmax}\parens{\frac{\mathcal{W}W_QW_K^\top\mathcal{W}^\top}{k}}\mathcal{W}W_V.
\end{equation}

The multi-head (MH) mechanism enables the network to jointly attend to different projections for different parts in the sequence, \textit{i.e.},
\begin{equation}
    \textrm{MH}\parens{\mathcal{W}} = \textrm{concat}\parens{\textrm{SA}_{enc, 1}\parens{\mathcal{W}}, \dots, \textrm{SA}_{enc, h}\parens{\mathcal{W}}}W_\textrm{concat},
\end{equation}
where $h$ is the number of heads, $W_\textrm{concat}$ is the set of trainable parameters associated with the concatenated representation, and each self-attention $i$ in the concatenation consists of its own set of trainable parameters $W_{Q, i}$, $W_{K, i}$, and $W_{V, i}$.

The transformer encoder then passes the MH output through two fully-connected (FC) layers. It repeats the entire block consisting of (SA--MH--FC) $N$ times and uses the residuals around each layer in the blocks during backpropagation. We denote the final encoded representation of the input sequence $\mathcal{W}$ as $F_\mathcal{W}$.

To meet the given constraints on the acting task $A$, intended emotion $E$, gender $G$, and handedness $H$ of the virtual agent, we append these variables to $F_\mathcal{W}$ and pass the combined representation through two fully-connected layers with trainable parameters $W_{FC}$ to obtain feature representations
\begin{equation}
    \bar{F_\mathcal{W}} = FC\parens{\begin{bmatrix} F_\mathcal{W}^\top & A^\top & E^\top & G^\top & H^\top \end{bmatrix}^\top; W_{FC}}.
\end{equation}

The transformer decoder operates similarly using the target sequence $\mathcal{Q}$, but with some important differences. First, it uses a masked multi-head (MMH) self-attention on the sequence, such that the attention for each element covers only those elements appearing before it in the sequence, \textit{i.e.},
\begin{equation}
    \textrm{MMH}\parens{\mathcal{Q}} = \textrm{concat}\parens{\textrm{SA}_{dec, 1}\parens{\mathcal{Q}}, \dots, \textrm{SA}_{dec, h}\parens{\mathcal{Q}}}W_\textrm{concat}.
\end{equation}
This ensures that the attention mechanism is causal and therefore usable at test time, when the full target sequence is not known apriori. Second, it uses the output of the MMH operation as the key and the value, and the encoded representation $\bar{F_\mathcal{W}}$ as the query, in an additional multi-head self-attention layer without any masking, \textit{i.e.},
\begin{equation}
    \resizebox{0.92\columnwidth}{!}{%
        $\textrm{MH}\parens{\bar{F_\mathcal{W}}, \mathcal{Q}} = \textrm{concat}\parens{\underbrace{\textrm{Att}_{dec, 1}\parens{\bar{F_\mathcal{W}}, \textrm{MMH}\parens{\mathcal{Q}}, \textrm{MMH}\parens{\mathcal{Q}}}, \dots}_{h \textrm{ entries}}}W_\textrm{concat}$.
    }
\end{equation}
It then passes the output of this multi-head self-attention through two fully-connected layers to complete the block. Thus, one block of the decoder is (SA--MMH--SA--MH--FC), and the transformer network uses $N$ such blocks. It also uses positional encoding of the target sequence upfront and uses the residuals around each layer in the blocks during backpropagation.

\section{Training the Transformer-Based Network}
Fig.~\ref{fig:network} shows the overall architecture of our transformer-based network. The word embedding layer transforms the words into feature vectors using the pre-trained GloVe model. The encoder and the decoder respectively consist of $N=2$ blocks of (SA--MH--FC) and (SA--MMH--SA--MH--FC). We use $h=2$ heads in the multi-head attention. The set of FC layers in each of the blocks maps to 200-dim outputs. At the output of the decoder, we normalize the predicted values so that they represent valid rotations. We train our network using the sum of three losses: the angle loss, the pose loss, and the affective loss. We compute these losses between the gesture sequences generated by our network and the original motion-captured sequences available as ground-truth in the training dataset.

\subsection{Angle Loss for Smooth Motions}
We denote the ground-truth relative rotation of each joint $j$ at time step $t$ as the unit quaternion $q_{j, t}$, and the corresponding rotation predicted by the network as $\hat{q}_{j, t}$. If needed, we correct $\hat{q}_{j, t}$ to have the same orientation as $q_{j, t}$. Then we measure the angle loss between each such pair of rotations as the squared difference of their Euler angle representations, modulo $\pi$. We use Euler angles rather than the quaternions in the loss function as it is straightforward to compute closeness between Euler angles using Euclidean distances. To ensure that the motions look smooth and natural, we also consider the squared difference between the derivatives of the ground-truth and the predicted rotations, computed at successive time steps. We write the net angle loss $\mathcal{L}_\textrm{ang}$ as
\begin{equation}
\begin{split}
    \mathcal{L}_\textrm{ang} =& \sum_t\sum_j\parens{ \textrm{Eul}\parens{q_{j, t}} - \textrm{Eul}\parens{\hat{q}_{j, t}}}^2 + \\ & \parens{\textrm{Eul}\parens{q_{j, t}} - \textrm{Eul}\parens{q_{j, t - 1}} - \textrm{Eul}\parens{\hat{q}_{j, t}} + \textrm{Eul}\parens{\hat{q}_{j, t - 1}}}^2.
\end{split}
\end{equation}

\subsection{Pose Loss for Joint Trajectories}
The angle loss only penalizes the absolute differences between the ground-truth and the predicted joint rotations and does not explicitly constrain the resulting poses to follow the same trajectory as the ground-truth at all time steps. To this end, we compute the squared norm difference between the ground-truth and the predicted joint positions at all time steps. Given the relative joint rotations and the offset $o_j$ of every joint $j$ from its parent, we can easily compute all the joint positions using forward kinematics (FK). Thus, we write the pose loss $\mathcal{L}_\textrm{pose}$ as
\begin{equation}
    \mathcal{L}_\textrm{pose} = \sum_t\sum_j \lVert \textrm{FK}\parens{q_{j, t}, o_j} - \textrm{FK}\parens{\hat{q}_{j, t}, o_j} \rVert^2.
\end{equation}

\begin{figure}[t]
    \centering
    \includegraphics[width=0.68\columnwidth]{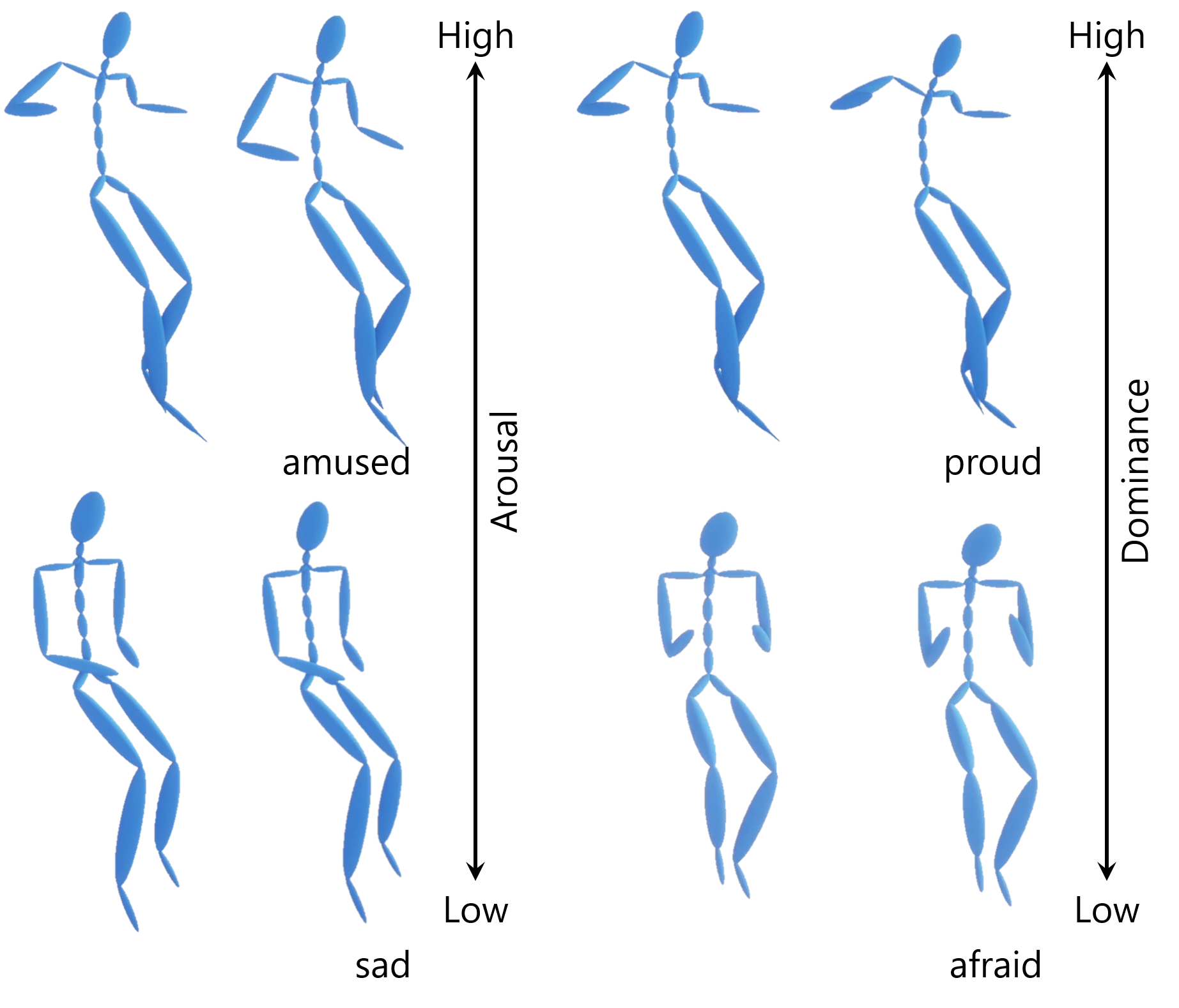}
    \caption{\textbf{Variance in emotive gestures.} Emotions with high arousal (\textit{e.g.}, amused) generally have rapid limb movements, while emotions with low arousal (\textit{e.g.}, sad) generally have slow and subtle limb movements. Emotions with high dominance (\textit{e.g.}, proud) generally have an expanded upper body and spread arms, while emotions with low dominance (\textit{e.g.}, afraid) have a contracted upper body and arms close to the body. Our algorithm uses these characteristics to generate the appropriate gestures.}
    \label{fig:arousal_dominance_gestures}
    \vspace{-15pt}
\end{figure}

\subsection{Affective Loss for Emotive Gestures}
To ensure that the generated gestures are emotionally expressive, we also penalize the loss between the gesture-based affective features of the ground-truth and the predicted poses. Prior studies in affective computing~\cite{effort_shape,affective_body_survey2,emotion_in_gesture} show that gesture-based affective features are good indicators of emotions that vary in arousal and dominance. Emotions with high dominance, such as pride, anger, and joy, tend to be expressed with an expanded upper body, spread arms, and upright head positions. Conversely, emotions with low dominance, such as fear and sadness, tend to be expressed with a contracted upper body, arms close to the body, and collapsed head positions. Again, emotions with high arousal, such as anger and amusement, tend to be expressed with rapid arm swings and head movements. By contrast, emotions with low arousal, such as relief and sadness, tend to be expressed with subtle, slow movements. Different valence levels are not generally associated with consistent differences in gestures, and humans often infer from other cues and the context. Fig.~\ref{fig:arousal_dominance_gestures} shows some gesture snapshots to visualize the variance of these affective features for different levels of arousal and dominance.

\begin{figure}[t]
    \centering
    \includegraphics[width=0.53\columnwidth]{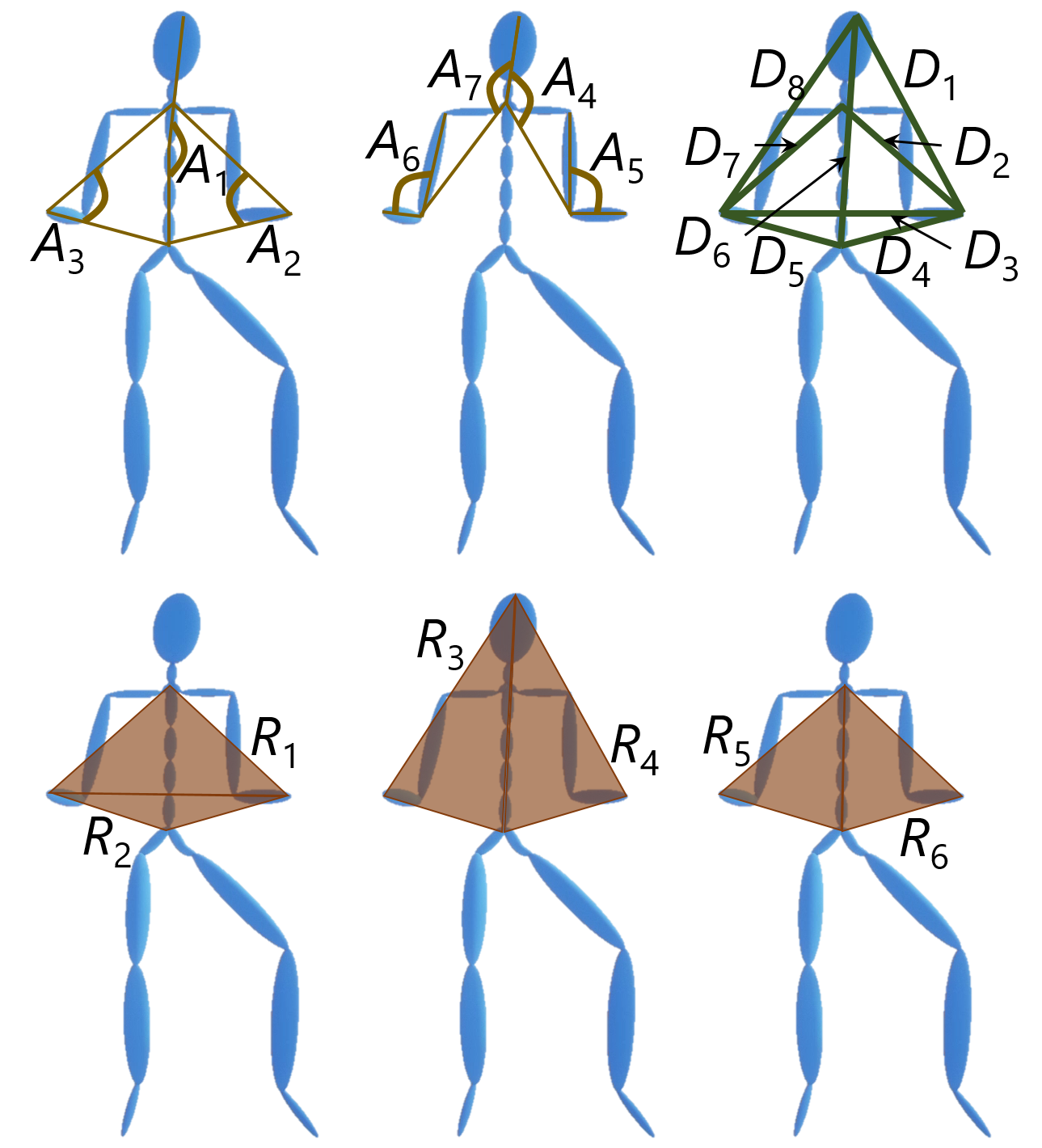}
    \caption{\textbf{Gesture-based affective features.} We use a total of 15 features: 7 angles, $A_1$ through $A_7$, 5 distance ratios, $\frac{D_1}{D_4}$, $\frac{D_2}{D_4}$, $\frac{D_8}{D_5}$, $\frac{D_7}{D_5}$, and $\frac{D_3}{D_6}$, and 3 area ratios, $\frac{R_1}{R_2}$, $\frac{R_3}{R_4}$, and $\frac{R_5}{R_6}$.}
    \label{fig:aff_features}
    \vspace{-15pt}
\end{figure}

We define scale-independent affective features using angles, distance ratios, and area ratios for training our network, following the same rationale as in~\cite{step}. Since, in our experiments, the virtual agent is sitting down, and only the upper body is expressive during the gesture sequences, only the joints at the root, neck, head, shoulders, elbows, and wrists move significantly. Therefore, we use these joints to compute our affective features. We show the complete list of affective features we use in Fig.~\ref{fig:aff_features}. Denoting the set of affective features computed from the ground-truth and the predicted poses at time $t$ as $a_t$ and $\hat{a}_t$ respectively, we write the affective loss $\mathcal{L}_\textrm{aff}$ as
\begin{equation}
    \mathcal{L}_\textrm{aff} = \sum_t \lVert a_t - \hat{a}^t \rVert^2.
\end{equation}

Combining all the individual loss terms, we write our training loss functions $\mathcal{L}$ as
\begin{equation}
    \mathcal{L} = \mathcal{L}_\textrm{ang} + \mathcal{L}_\textrm{pose} + \mathcal{L}_\textrm{aff} + \lambda\lVert W \rVert,
    \label{eq:total_loss}
\end{equation}
where $W$ denotes the set of all trainable parameters in the full network, and $\lambda$ is the regularization factor.

\section{Results}\label{sec:results}
This section elaborates on the database we use to train, validate, and test our method. We also report our training routine, the performance of our method compared to the ground-truth, and the current \sota~method for generating gestures aligned with text input. We also perform ablation studies to show the benefits of each of the components in our loss function: the angle loss, the pose loss, and the affective loss.

\subsection{Data for Training, Validation and Testing}
We evaluate our method on the MPI emotional body expressions database~\cite{EBEDB}. This database consists of 1,447 motion-captured sequences of human participants performing one of three acting tasks: narrating a sentence from a story, gesticulating a scenario given as a sentence, or gesticulating while speaking a line in a conversation. Each sequence corresponds to one text sentence and the associated gestures. For each sequence, the following annotations of the intended emotion $E$, gender $G$, and handedness $H$, are available:
\begin{itemize}
    \item $E$ as the VAD representation for one of ``afraid'', ``amused'', ``angry'', ``ashamed'', ``disgusted'', ``joyous'', ``neutral'', ``proud'', ``relieved'', ``sad'', or ``surprised'',
    \item $G$ is either female or male, and
    \item $H$ is either left or right.
\end{itemize}
Each sequence is captured at 120 fps and is between 4 and 20 seconds long. We pad all the sequences with our EoS pose (Sec.~\ref{subsec:ges_rep}) so that all the sequences are of equal length. Since the sequences freeze at the end of the corresponding sentences, padding with the EoS pose often introduces small jumps in the joint positions and the corresponding relative rotations when any gesture sequence ends. To this end, we have designed our training loss function (Eq.~\ref{eq:total_loss}) to ensure smoothness and generate gestures that transition smoothly to the EoS pose after the end of the sentence.

\subsection{Training and Evaluation Routines}
We train our network using the Adam optimizer~\cite{adam} with a learning rate of 0.001 and a weight decay of 0.999 at every epoch. We train our network for 600 epochs, using a stochastic batch size of 16 without replacement in every iteration. We have a total of 26,264,145 trainable parameters in our network. We use 80\% of the data for training, validate the performance on 10\% of the data, and test on the remaining 10\% of the data. The total training takes around 8 hours using an Nvidia GeForce GTX 1080Ti GPU. At the time of evaluation, we initialize the transformer decoder with $T=20$ (Fig.~\ref{fig:network}) time steps of the SoS pose and keep using the past $T=20$ time steps to generate the gesture at every time step.

\subsection{Comparative Performance}\label{subsec:performance_eval}
We compare the performance of our network with the transformer-based text-to-gesture generation network of Yoon et al.~\cite{cospeech_gestures} because this method is the closest to our work. To make a fair comparison, we perform the following as per their original paper:
\begin{itemize}
    \item use the eight upper body joints (three each on the two arms, neck, and head) for their method,
    \item use PCA to reduce the eight upper body joints to 10-dimensional features,
    \item retrain their network on the MPI emotional body expressions database~\cite{EBEDB}, using the same data split as in our method, and the hyperparameters provided by the authors,
    \item compare the performances only on the eight upper body joints.
\end{itemize}

\begin{table}[t]
    \centering
    \caption{\textbf{Mean pose errors.} For each listed method, this is the mean Euclidean distance of all the joints over all the time steps from all the ground-truth sequences over the entire test set. The mean error for each sequence is computed relative to the mean length of the longest diagonal of the 3D bounding box of the virtual agent in that sequence.}
    \label{tab:mean_pose_errors}
    \begin{tabular}{lc}
        \toprule
        Method & Mean pose error \\
        \midrule
        Yoon et al.~\cite{cospeech_gestures} & 1.57 \\
        Our method, no angle loss & 0.07 \\
        Our method, no pose loss & 0.06 \\
        Our method, no affective loss & 0.06 \\
        Our method, all losses & 0.05 \\
        \bottomrule
    \end{tabular}
    \vspace{-15pt}
\end{table}

\begin{figure}[t]
    \centering
    \includegraphics[width=0.75\columnwidth]{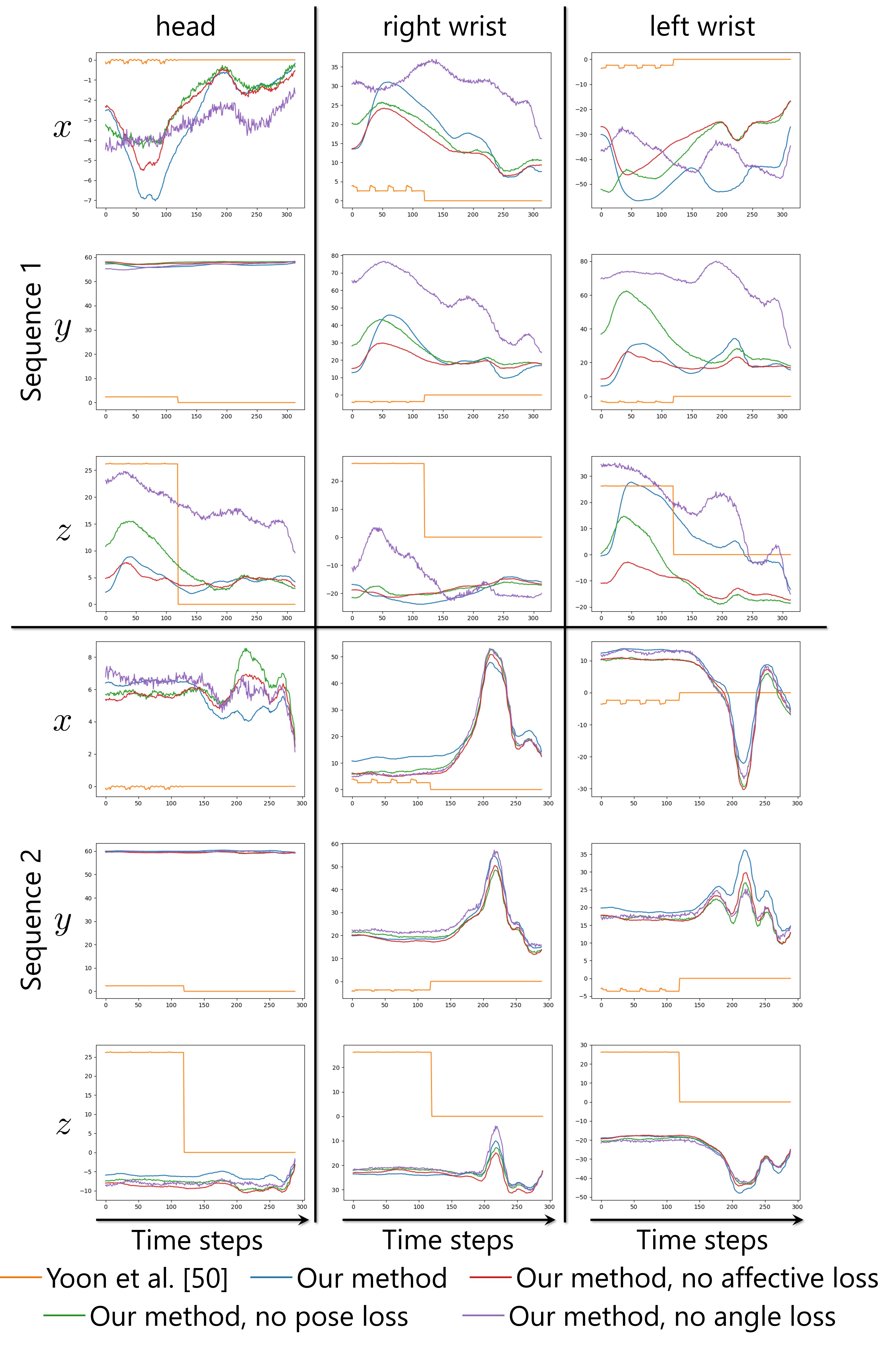}
    \caption{\textbf{End-effector trajectories.} The trajectories in the three coordinate directions for the head and two wrists. We show two sample sequences from the test set, as generated by all the methods. Removing the angle loss makes the trajectory heavily jerky. Removing the pose loss makes our method unable to follow the desired trajectory. Removing the affective loss reduces the variations corresponding to emotional expressiveness. Yoon et al.'s method~\cite{cospeech_gestures} is unable to generate large amplitude variations in the trajectories because it works with a dimension-reduced representation of the sequences.}
    \label{fig:trajs}
    \vspace{-15pt}
\end{figure}

We report the mean pose error from the ground-truth sequences over the entire held-out test set for both Yoon et al.~\cite{cospeech_gestures} and our method in Table~\ref{tab:mean_pose_errors}. For each test sequence and each method, we compute the total pose error for all the joints at each time step and calculate the mean of these errors across all time steps. We then divide the mean error by the mean length of the longest diagonal of the 3D bounding box of the virtual agent to get the normalized mean error. To obtain the mean pose error for the entire test set, we compute the mean of the normalized mean errors for all the test sequences. We also plot the trajectories of the three end-effector joints in the upper body, head, left wrist, and right wrist, independently in the three coordinate directions, for two diverse sample sequences from the test set in Fig.~\ref{fig:trajs}. We ensure diversity in the samples by choosing a different combination of the gender, handedness, acting task, and intended emotion of the gesture for each sample. 

We observe from Table~\ref{tab:mean_pose_errors} that our method reduces the mean pose error by around 97\% over Yoon et al.~\cite{cospeech_gestures}. From the plots in Fig.~\ref{fig:trajs}, we can observe that unlike our method, Yoon et al.'s method is unable to generate the high amplitude oscillations in motion, leading to larger pose errors. This is because their lower-dimensional representation of pose motions does not sufficiently capture the oscillations. Moreover, the gestures generated by Yoon et al.’s method did not produce any movements in the $z$-axis. Instead, they confined the movements to a particular $z$-plane. The step in their method in the $z$-axis occurs when the gesture returns to the EoS rest pose, which is in a different $z$-plane.

\begin{figure}[t]
    \centering
    \includegraphics[width=0.8\columnwidth]{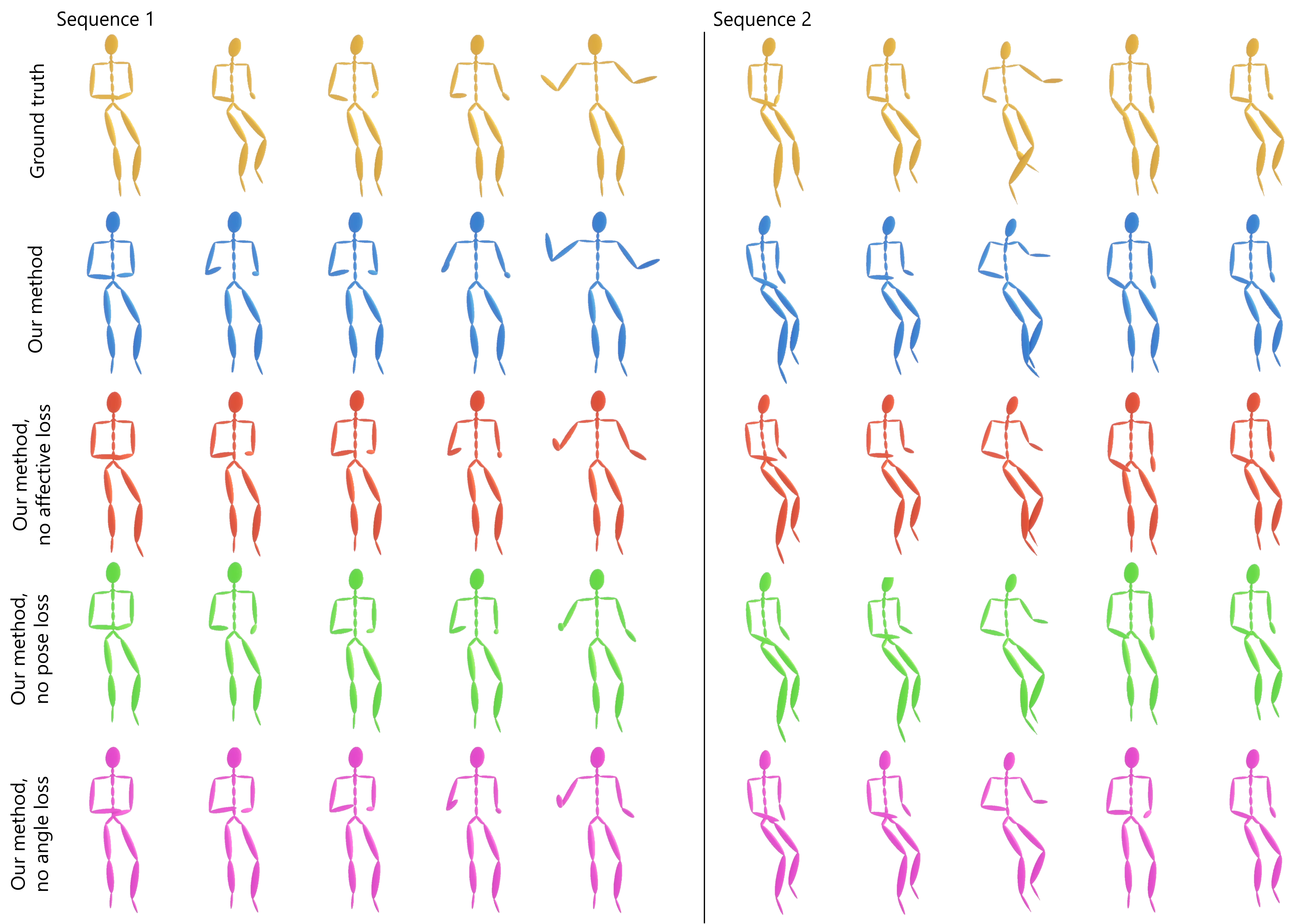}
    \caption{\textbf{Ablation studies.} Snapshots of gestures at five time steps from two sample ground-truth sequences in the test set, and the gestures at the same five time steps as generated by our method and its different ablated versions. The full sequences of these gestures are available in our supplementary video.}
    \label{fig:ablation}
    \vspace{-15pt}
\end{figure}

\subsection{Ablation Studies}\label{subsec:ablation}
We compare the performance between different ablated versions of our method. We test the contribution of each of the three loss terms, angle loss, pose loss, and affective loss, in Eq.~\ref{eq:total_loss} by removing them from the total loss one at a time and training our network from scratch with the remaining losses. Each of these ablated versions has a higher mean pose error over the entire test set than our actual method, as we report in Table~\ref{tab:mean_pose_errors}. To visualize the performance differences, we show in Fig.~\ref{fig:trajs} sample end-effector trajectories in the same setup as described in Sec.~\ref{subsec:performance_eval}. We also show snapshots from the two sample gesture sequences generated by all the ablated versions in Fig.~\ref{fig:ablation}. We show the full gesture sequences of these and other samples in our supplementary video.

We can observe from Fig.~\ref{fig:trajs} that the gestures become heavily jerky without the angle loss. When we add in the angle loss but remove the pose loss, the gestures become smoother but still have some jerkiness. This shows that the pose loss also lends some robustness to the generation process. The other major drawback in removing either the angle or the pose loss is that the network can only change the gesture between time steps within some small bounds, making the overall animation sequence appear rigid and constricted.

When we remove only the affective loss from Eq.~\ref{eq:total_loss}, the network can generate a wide range of gestures, leading to animations that appear fluid and plausible. However, the emotional expressions in the gestures, such as spreading and contracting the arms and shaking the head, are not consistent with the intended emotions.

\subsection{Interfacing the VR Environment}
Given a sentence of text, we can generate the gesture animation files at an interactive rate of 3.2 ms per frame, or 312.5 frames per second, on average on an Nvidia GeForce GTX 1080Ti GPU.

We use gender and handedness to determine the virtual agent's physical attributes during the generation of gestures. Gender impacts the pose structure. The handedness determines the hand for one-handed or longitudinally asymmetrical gestures. To create the virtual agents, we use low-poly humanoid meshes with no textures on the face. We use the pre-defined set of male and female skeletons in the MPI emotional body motion database~\cite{EBEDB} for the gesture animations. We assign a different model to each of these skeletons, matching their genders. We manually correct any visual distortions caused by a shape mismatch between the pre-defined skeletons and the low-poly meshes.

We use Blender 2.7 to rig the generated animations to the humanoid meshes. To ensure a proper rig, we modify the rest pose of the humanoid meshes to match the rest pose of our pre-defined skeletons. To make the meshes appear more life-like, we add periodic blinking and breathing movements to the generated animations using blendshapes in Blender.

We prepare our VR environment using Unreal 4.25. We place the virtual agents on a chair in the center of the scene in full focus. The users can interact with the agent in two ways. They can either select a story that the agent narrates line by line using appropriate body gestures or send lines of text as part of a conversation to which the agent responds using text and associated body gestures. We show the full demos in our supplementary video. We use synthetic, neutral-toned audio aligned with all our generated gestures to understand the timing of the gestures with the text. However, we do not add any facial features or emotions in the audio for the agents since they are dominant modalities of emotional expression and make a fair evaluation of the emotional expressiveness of the gestures difficult. For example, if the intended emotion is happy, and the agent has a smiling face, observers are more likely to respond favorably to any gesture with high valence or arousal.

\section{User Study}\label{sec:user_study}
We conduct a web-based user study to test two major aspects of our method: the correlation between the intended and the perceived emotions of and from the gestures, and the quality of the animations compared to the original motion-captured sequences.

\subsection{Procedure}
The study consisted of two sections and was about ten minutes long. In the first section, we showed the participant six clips of virtual agents sitting on a chair and performing randomly selected gesture sequences generated by our method, one after the other. We then asked the participant to report the perceived emotion as one of multiple choices. Based on our pilot study, we understood that asking participants to choose from one of 11 categorical emotions in the EBEDB dataset~\cite{EBEDB} was overwhelming, especially since some of the emotion terms were close to each other in the VAD space (\textit{e.g.}, joyous and amused). Therefore, we opted for fewer choices to make it easier for the participants and reduce the probability of having too many emotion terms with similar VAD values in the choices. For each sequence, we, therefore, provided the participant with four choices for the perceived emotion. One of the choices was the intended emotion, and the remaining three were randomly selected. For each animation, randomly choosing three choices can unintentionally bias the participant's response (for instance, if the intended emotion is ``sad'' and the random options are ``joyous'', ``amused'' and ``proud''). However, the probability of such a set of choices drops exponentially as we consider multiple sequences for each participant and multiple participants in the overall study.

\begin{table}[t]
    \centering
    \caption{\textbf{Likert scale markers to asses quality of gestures}. We use the following markers in our five-point Likert scale}
    \label{tab:likert_scale}
    \resizebox{\columnwidth}{!}{%
    \begin{tabular}{ll}
        \toprule
        Very Unnatural & \textit{e.g.}, broken arms or legs, torso at an impossible angle \\
        Not Realistic & \textit{e.g.}, limbs going inside the body or through the chair \\
        Looks OK & No serious problems, but does not look very appealing \\
        Looks good & No problems and the gestures look natural \\
        Looks great! & The gestures look like they could be from a real person \\
    \bottomrule
    \end{tabular}
    }
    \vspace{-15pt}
\end{table}

In the second section, we showed the participant three clips of virtual agents sitting on a chair and performing a randomly selected original motion-captured sequence and three clips of virtual agents performing a randomly selected generated gesture sequence, one after the other. We showed the participant these six sequences in random order. We did not tell the participant which sequences were from the original motion-capture and which sequences were generated by our method. We asked the participant to report the naturalness of the gestures in each of these sequences on a five-point Likert scale, consisting of the markers mentioned in Table~\ref{tab:likert_scale}.

We had a total of 145 clips of generated gestures and 145 clips of the corresponding motion-captured gestures. For every participant, we chose all the 12 random clips across the two sections without replacement. We did not notify the participant apriori which clips had motion-captured gestures and which clips had our generated gestures. Moreover, we ensured that in the second section, none of the three selected generated gestures corresponded to the three selected motion-captured gestures. Thus, all the clips each participant looked at were distinct. However, we did repeat clips at random across participants to get multiple responses for each clip.

\subsection{Participants}
Fifty participants participated in our study, recruited via web advertisements. To study the demographic diversity, we asked the participants to report their gender and age group. Based on the statistics, we had 16 male and 11 female participants in the age group of 18-24, 15 male and seven female participants in the age group of 25-34, and one participant older than 35 who preferred not to disclose their gender. However, we did not observe any particular pattern of responses based on the demographics.

\subsection{Evaluation}
We analyze the correlation between the intended and the perceived emotions from the first section of the user study and the reported quality of the animations from the second section. We also summarize miscellaneous user feedback.

\subsubsection{Correlation between Intended and Perceived Emotions}
\begin{figure}[t]
    \centering
    \includegraphics[width=0.8\columnwidth]{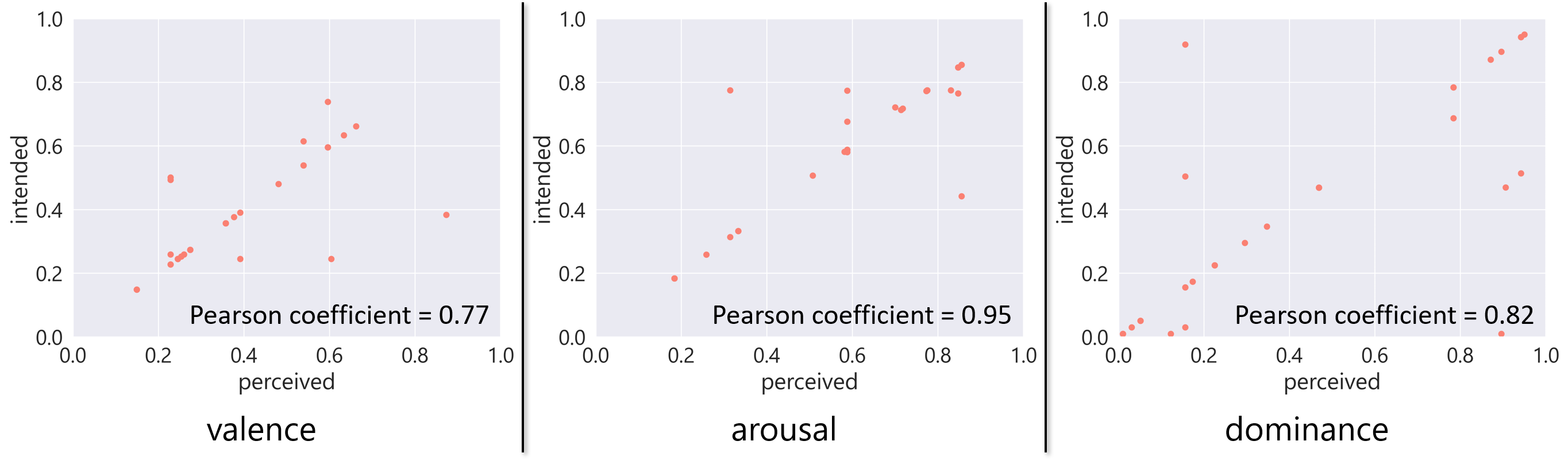}
    \caption{\textbf{Valence, arousal, and dominance distributions.} Distribution of values from the intended and perceived emotions in the valence, arousal, and dominance dimensions for gestures in the study. All the distributions indicate strong positive correlation between the intended and the perceived values, with the highest correlation in arousal and the lowest in valence.}
    \label{fig:vad_dist}
    \vspace{-15pt}
\end{figure}

Each participant responded to six random sequences in the first section of the study, leading to a total of 300 responses. We convert the categorical emotion terms from these responses to the VAD space using the mapping of NRC-VAD~\cite{nrc_vad}. We show the distribution of the valence, arousal, and dominance values of the intended and perceived emotions in Fig.~\ref{fig:vad_dist}.

We compute the Pearson correlation coefficient between the intended and perceived values in each of the valence, arousal, and dominance dimensions. A Pearson coefficient of 1 indicates maximum positive linear correlation, 0 indicates no correlation, and -1 indicates maximum negative linear correlation. In practice, any coefficient larger than 0.5 indicates a strong positive linear correlation. We hypothesize that intended and the perceived values in all three dimensions have such a strong positive correlation.

We observe a Pearson coefficient of 0.77, 0.95, and 0.82, respectively, between the intended and the perceived values in the valence, arousal, and dominance dimensions. Thus, the values in all three dimensions are strongly positively correlated, satisfying our hypothesis. The values also indicate that the correlation is stronger in the arousal and the dominance dimensions and comparatively weaker in the valence dimension. This is in line with prior studies in affective computing~\cite{effort_shape,affective_body_survey2}, which show that humans can consistently perceive arousal and dominance from gesture-based body expressions.

\subsubsection{Quality of Gesture Animations}
\begin{figure}[t]
    \centering
    \includegraphics[width=0.7\columnwidth]{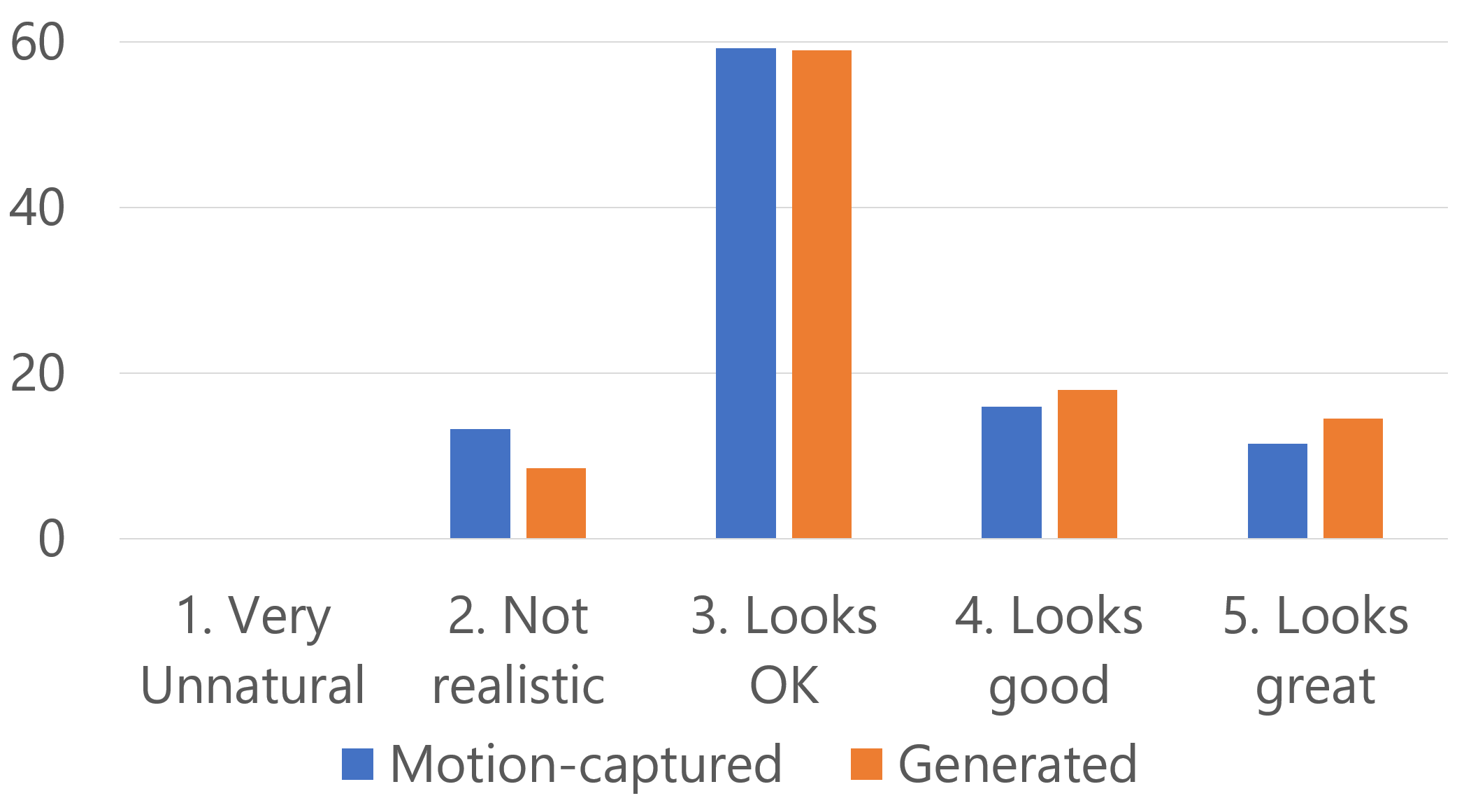}
    \caption{\textbf{Responses on the Quality of Gestures.} A small fraction of participants responded to the few gesture sequences that had some stray self-collisions, and therefore found these sequences to not be realistic. The vast majority of the participants found both the motion-captured and generated gestures to look OK (plausible) on the virtual agents. A marginally higher percentage of participants reported that our generated gesture sequences looked better on the virtual agents that the original motion-captured gesture sequences.}
    \label{fig:quality_responses}
    \vspace{-15pt}
\end{figure}

Each participant responded to three random motion-captured and three randomly generated sequences in the second section of the study. Therefore, we have a total of 150 responses on both the motion-captured and the generated sequences. We summarize the percentage of responses of each of the five points in the Likert scale in Fig.~\ref{fig:quality_responses}. We consider a minimum score of 3 on our Likert scale to indicate that the participant found the corresponding gesture plausible. By this criterion, we observe that 86.67\% of the responses indicated the virtual agents performing the motion-captured sequences have plausible gestures and 91.33\% of the responses the virtual agents performing the generated sequences have plausible gestures. In fact, we observe that a marginally higher percentage of responses scored the generated gestures 4 and 5 (2.00\% and 3.33\% respectively), compared to the percentage of responses with the same score for the motion-captured gestures. This, coupled with the fact that participants did not know apriori which sequences were motion-captured and generated, indicates that our generated sequences were perceived to be as realistic as the original motion-captured sequences. One possible explanation of participants rating our generated gestures marginally more plausible than the motion-captured gestures is that our generated poses return smoothly to a rest pose after the end of the sentence. The motion-captured gestures, on the other hand, freeze at the end-of-the-sentence pose.

\subsubsection{Miscellaneous Feedback}
Our virtual agents only express emotions through gestures and do not use any other modalities such as faces or voices. Therefore, we expected some participants taking the study to be distracted by the lack of emotions on the face or to be unable to determine the emotions based only on the gestures, without supporting cues from the other modalities. Indeed, 14\% of the participants reported they were distracted by the lack of facial emotions, 10\% were unable to determine the emotions based on only the gestures, and 8\% experienced both difficulties.
% Therefore, adding emotive facial expressions and synchronizing them with the text and the gesture expressions is an important design aspect we will consider for our system.

\section{Conclusion}\label{sec:conclusion}
We present a novel method that takes in natural language text one sentence at a time and generates 3D pose sequences for virtual agents corresponding to emotive gestures aligned with that text. Our generative method also considers the intended acting task of narration or conversation, the intended emotion based on the text and the context, and the intended gender and handedness of the virtual agents to generate plausible gestures. We can generate these gestures in a few milliseconds on an Nvidia GeForce GTX 1080Ti GPU. We also conducted a web study to evaluate the naturalness and emotional expressiveness of our generated gestures. Based on the 600 total responses from 50 participants, we found a strong positive correlation between the intended emotions of the virtual agents' gestures and the emotions perceived from them by the respondents, with a minimum Pearson coefficient of 0.77 in the valence dimension. Moreover, around 91\% of the respondents found our generated gestures to be at least plausible on a five-point Likert Scale.

\section{Limitations and Future Work}\label{sec:lim_fw}
Our work has some limitations. First, we train our network to learn mappings from complete text sentences to gestures. We can improve this by exploring a more granular phrase-level mapping from text to gestures to gain insights on how gestures corresponding to parts of sentences can be combined to produce gestures for full sentences. Second, our generated gestures return to the EoS pose after every gesticulating every sentence. This is because all the samples in the EBEDB dataset~\cite{EBEDB} start from a rest pose. As a result, we cannot exploit any information related to the continuity between gestures that correspond to adjacent sentences. A simple method to extend this approach is to use the last window of the current sentence as an initialization for the next sentence. However, without any ground-truth information on the continuity between gesture, it is difficult to train or evaluate the transitioning gestures. As part of our future work, we plan to explore other techniques to enforce such continuity. Third, We only consider the VAD representation for the categorical emotion terms associated with the texts. This simplifies our network design and the evaluation of the emotions perceived by the participants in our study. In the future, we plan to explore the correlations between the VAD representations of words in the text and the associated categorical emotions. We also plan to study the interrelations of these VAD representations with the gender, age, and ethnicity of the subjects, to build more sophisticated maps from texts to a more diverse range of emotive gestures. We would also like to integrate our emotive gesture generation algorithm with social VR systems and use them for socially-aware conversational agents. 

Lastly, we only consider text-based emotive gesture generation, but no facial expression or expressive voice tones. In real-world scenarios, facial expressions and voice tones tend to play dominant roles in conveying the emotions and may occupy the user's focus. Consequently, in our current video results and studies, we evaluated the effectiveness of our gesture generation approach without any using facial or vocal expressions, which is similar to other methods for evaluating gestures~\cite{trimodal_gesture_generation,gesticulator}. This way, we ensure that the user mainly focuses on emotive gestures. As part of future work, it would be useful to combine our work with varying facial expressions corresponding to different emotions and vary the emotional tone in voices. Furthermore, we would like to evaluate the relative benefits of combining different modalities, such as emotive gestures, facial expressions, and voice tones.


\clearpage
\bibliographystyle{abbrv-doi}

\bibliography{main}
\end{document}